\documentclass{article}

\usepackage{arxiv}

\usepackage[utf8]{inputenc} % allow utf-8 input
\usepackage[T1]{fontenc}    % use 8-bit T1 fonts
\usepackage{hyperref}       % hyperlinks
\usepackage{url}            % simple URL typesetting
\usepackage{booktabs}       % professional-quality tables
\usepackage{amsfonts}       % blackboard math symbols
\usepackage{amsmath}
\usepackage{nicefrac}       % compact symbols for 1/2, etc.
\usepackage{microtype}      % microtypography
\usepackage{graphicx}

\def\apj{Astrophys.\ J.\ }

\def\apjs{Astrophys.\ J.\ Suppl.\ }

\def\aap{Astron.\ Astrophys.\ }

\def\nat{Nature}

\def\prd{Phys.\ Rev.\ D}
\def\app{Astropart.\ Phys.\ }
\def\plb{Phys.\ Lett.\ B}
\def\jcap{JCAP}

\def\na{Nat.\ Astron.\ }

\def\cpc{Chin.\ Phys.\ C}

\title{ Novel pre-burst stage of gamma-ray bursts from machine learning\thanks{Published in J.~High Energy Astrophys. 32 (2021) 78-86, \url{https://doi.org/10.1016/j.jheap.2021.09.002}} }

\author{
  Yingtian Chen \\
  School of Physics, Peking University, Beijing 100871, China \\
  Department of Astronomy, University of Michigan, Ann Arbor, MI 48109, USA \\
   \And
  Bo-Qiang Ma \thanks{Corresponding author at: School of Physics, Peking University, Beijing 100871, China. Email address: mabq@pku.edu.cn} \\
  School of Physics, Peking University, Beijing 100871, China \\
  State Key Laboratory of Nuclear Physics and Technology, Peking University, Beijing 100871, China \\
  Center for High Energy Physics, Peking University, Beijing 100871, China \\
  Collaborative Innovation Center of Quantum Matter, Beijing, China \\
}

\begin{document}
\maketitle

\begin{abstract}
Gamma-ray bursts~(GRBs), as extremely energetic explosions in the universe, are widely believed to consist of two stages: the prompt phase and the subsequent afterglow. Recent studies indicate that some high-energy photons are emitted earlier at source than the prompt phase. Due to the light speed variation, these high-energy photons travel more slowly than the low-energy photons, so that they are observed after the prompt low-energy photons at the detector. Based on the data from the Fermi Gamma-ray Space Telescope~(FGST), we analyse the photon distribution before the prompt emission in detail and propose the existence of a hitherto unknown pre-burst stage of GRBs by adopting a classification method of machine learning. Analysis on the photons automatically selected by machine learning also produces a light speed variation at $E_{\mathrm{LV}}=\mathrm{3.55\times 10^{17}~GeV}$.
\end{abstract}

% keywords can be removed

\section{Introduction} \label{sec:1}
Gamma-ray bursts~(GRBs) were first discovered by the Vela satellites in 1967~\cite{Vela}. The random occurrence of GRBs and the limitation of positioning capability of the earlier facilities limited the progress of studying these luminous explosions. The situation has changed dramatically since the launch of the Fermi Gamma-ray Space Telescope~(FGST)~\cite{GBM,LAT}, with the detection of over 1600 GRBs including 25 bright ones with known redshifts~\cite{Ackermann:2013zfa,Bissaldi:2015mfa}.
FGST has two instruments, the Gamma-ray Burst Monitor~(GBM)~\cite{GBM}, which is sensitive to X-rays and gamma photons with observed energies between $\mathrm{8~keV}$ and $\mathrm{40~MeV}$, and the Large Area Telescope~(LAT)~\cite{LAT}, which detects higher-energy photons from $\mathrm{20~MeV}$ to more than $\mathrm{300~GeV}$. The combination of LAT and GBM gives a wide energy range of GRB photon detection.
GBM is triggered when it detects a sudden gamma-ray enhancement, which is usually regarded as the precursor of a GRB. Previous studies focused on the events observed after the trigger time, that is, the prompt phase and the subsequent afterglow~(in this paper, we combine them as the ``main-burst'' stage). The prompt phase lasts commonly several tens of seconds with the release of a large content of total energy~($\mathrm{10^{51}~ergs}$). Following the prompt phase is the afterglow, which lasts from few weeks to a year, offering a method to measure the redshifts of GRBs and their extragalactic origin.

Recent studies~\cite{shaolijing,zhangshu,Xu:2016zxi,Xu:2016zsa,note-added,Xu:2018,Liu:2018} on the light speed variation from GRBs suggested that some high-energy photons, though observed after the onset of the prompt phase, are emitted before the main-burst at source. We simply call these earlier emission as the ``pre-burst'' stage of GRBs.
In this work, we statistically analyse the data from FGST of 25 bright GRBs with measured redshifts~\cite{LATdata,redshift_data}
by adopting a classification method of machine learning.
The machine learning method automatically provides 4 main classifications to suggest the existence of the pre-burst stage of GRBs in all cases.
Analysis on the photons automatically classified by machine learning also produce a light speed variation at $E_{\mathrm{LV}}=\mathrm{3.55\times 10^{17}~GeV}$.

\section{Results} \label{sec:2}
\subsection{A first glimpse on the FGST data}\label{sec:2.1}
It is suggested in References~\cite{Xu:2016zxi,Xu:2016zsa,Xu:2018,Liu:2018} that the first main peak of the GBM light curve can be regarded as the characteristic time for the low energy photons of the prompt phase~(see Table~\ref{tab:1}). We set the first main peak time $t^{\mathrm{peak}}$ as time zero $t^0=0$ in our analysis.

Considering the redshift effect of cosmic scale, we can convert the observed time~($t_{\mathrm{obs}}$) and energy~($E_{\mathrm{obs}}$) of each photon event to the re-scaled time~$\tau$ and the intrinsic energy $E_{\mathrm{in}}$) as
\begin{equation}
\begin{aligned}
\tau &=t_{\mathrm{obs}}/(1+z), \\
E_{\mathrm{in}} &=(1+z)E_{\mathrm{obs}},
\label{eq:1}
\end{aligned}
\end{equation}
where $z$ is the redshift of each GRB~(see Table~\ref{tab:1}).
Unlike previous studies, we also focus on the photons observed before time zero~(in fact we should focus on photons emitted at source before time zero. But it is impossible to clearly identify these photons without considering the time difference caused by the light speed variation at present. This problem will be discussed later after the consideration of the light speed variation).

According to the default values of the LAT website~\cite{LATdata}, we select photons from all 25 GRBs with known redshifts~(see Table~\ref{tab:1}) within an time window of $\mathrm{-100~s}<\tau<\mathrm{100~s}$, an energy interval of $\mathrm{100~MeV}<E_{\mathrm{in}}<\mathrm{300~GeV}$ and a search radius range of $\theta<\mathrm{15^\circ}$.

\begin{table}
	\caption{25 GRBs analysed in this work}
	\centering
	\begin{tabular}{cccccc}
		\toprule
		GRB & $z^a$ & $t_{\mathrm{obs}}^{0}\mathrm{(s)}^b$ & GRB & $z^a$ & $t_{\mathrm{obs}}^{0}\mathrm{(s)}^b$ \\
		\midrule
		160625B & 1.406 & 0.284 & 120624A & 2.1974 & 8.314 \\
		160509A & 1.17 & 13.920 & 110731A & 2.83 & 0.488 \\
		150514A & 0.807 & 0.958 & 100728A & 1.567 & 54.004 \\
		150403A & 2.06 & 11.388 & 100414A & 1.368 & 0.288 \\
		150314A & 1.758 & 1.504 & 091208B & 1.063 & 0.722 \\
		141028A & 2.33 & 13.248 & 091003A & 0.8969 & 5.410 \\
		131231A & 0.642 & 23.040 & 090926A & 2.1071 & 4.320 \\
		131108A & 2.40 & 0.128 & 090902B & 1.822 & 9.768 \\
		130702A & 0.145 & 1.788 & 090510A & 0.903 & -0.032 \\
		130518A & 2.488 & 25.854 & 090328A & 0.736 & 5.378 \\
		130427A & 0.3399 & 0.544 & 090323A & 3.57 & 15.998 \\
		120729A & 0.80 & 1.488 & 080916C & 4.35 & 5.984 \\
		120711A & 1.405 & 69.638 &  &  & \\
		\bottomrule
	\end{tabular} \\
	\footnotesize{$a$. The redshift data can be found in Ref.~\cite{redshift_data}.} \\
	\footnotesize{$b$. The time zeros~(observed times of the first main peak) are from Ref.~\cite{Xu:2018}.} \\
	\label{tab:1}
\end{table}

Since photons with energy greater than the threshold of 1 GeV is widely regarded as high energy photons~\cite{Abdo:1, Abdo:2}, we adopt the threshold of 1 GeV to identify high energy photons. Considering different distributions between higher-energy and lower-energy photons, we divide all photons into two groups according to their energy ranges: photons with $E_{\mathrm{in}}<1~\mathrm{GeV}$ are assigned to Group~1 and photons with $E_{\mathrm{in}}>1~\mathrm{GeV}$ are assigned to Group~2.

It is certain that the spatial distribution of photons is not random in the time interval of $\tau > 0$. We note that such non-randomness also exists in the time interval of $\tau < 0$: photons of each GRB are distributed more densely near a certain center.
We use the average coordinate of photons as the center of each GRB. This method is reasonable when the search radius is not big~($\theta<\mathrm{15^\circ}$).

\begin{figure}
	\begin{center}
		\includegraphics[width=0.45\linewidth]{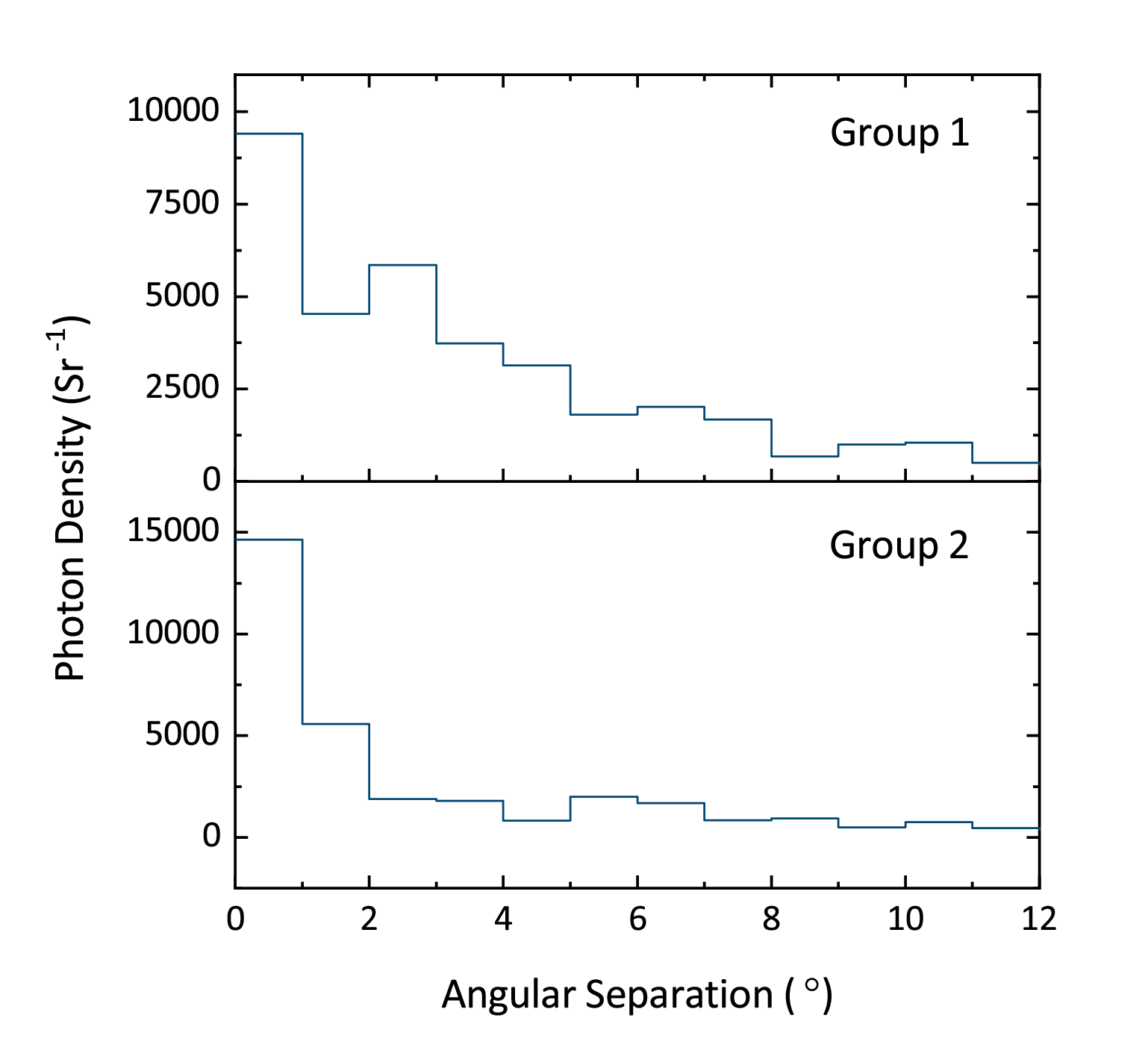}
		\caption{{The density-angle relation of GRB photon events with energies $E_{\mathrm{in}}<1~\mathrm{GeV}$ (Group~1) and $E_{\mathrm{in}}>1~\mathrm{GeV}$ (Group~2).} The horizontal coordinates are the angular separation of the photons to the centers.}
		\label{fig:1}
	\end{center}
\end{figure}

By plotting the relation between the photon density and the angular separation~(see Figure~\ref{fig:1}), we can demonstrate such non-randomness concretely. It can be seen that the photon density decreases rapidly with the increase of the angular separation, especially for the higher-energy Group~2.

\begin{figure}
	\begin{center}
		\includegraphics[width=0.45\linewidth]{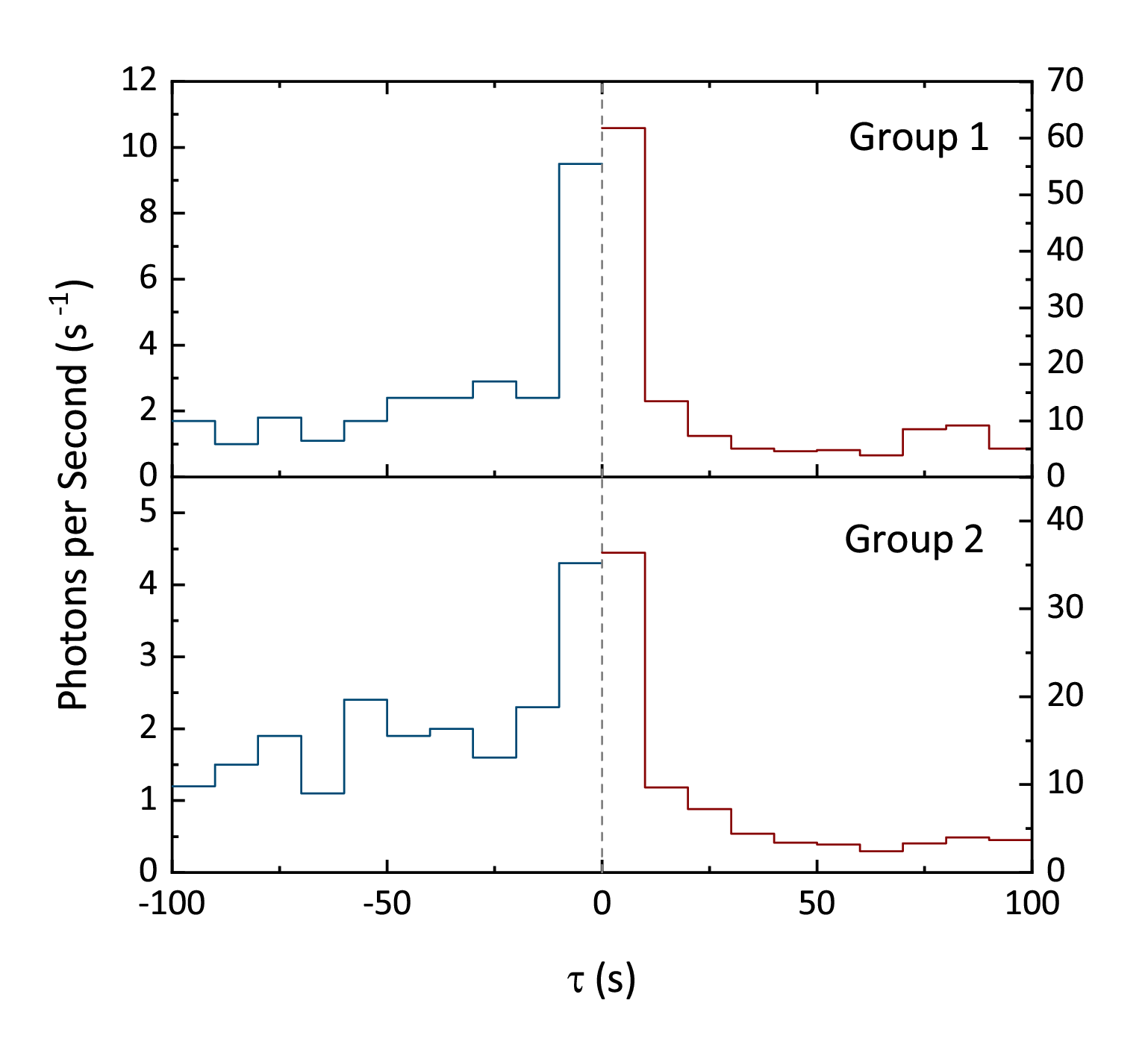}
		\caption{{The photon density-time relation of GRB photon events.} Since the scale of the blue curve on the left of each figure is quite different from the scale of the red curve on the right, we adopt different vertical ordinates.}
		\label{fig:2}
	\end{center}
\end{figure}

The time distribution of photons is also not random in the time interval of $\tau < 0$. This relation can be directly shown by plotting the light curve~(see Figure~\ref{fig:2}). It can be seen that as time approaches time zero from both the positive and negative axes, the photon density increases gradually.


Figures~\ref{fig:1} and \ref{fig:2} show that even before the main-burst, the photon distribution still has certain convergence.Such convergence is consistent for both low and high energy photons from Groups~1 and 2 respectively. It is reasonable to conclude that some photons observed before the main-burst still belong to the GRB rather than the background. Although the above discussion does not consider the time difference caused by the light speed variation, the aforementioned fact implies that there might be a pre-burst stage before the main-burst stage at source.

\begin{figure}
	\begin{center}
		\includegraphics[width=0.45\linewidth]{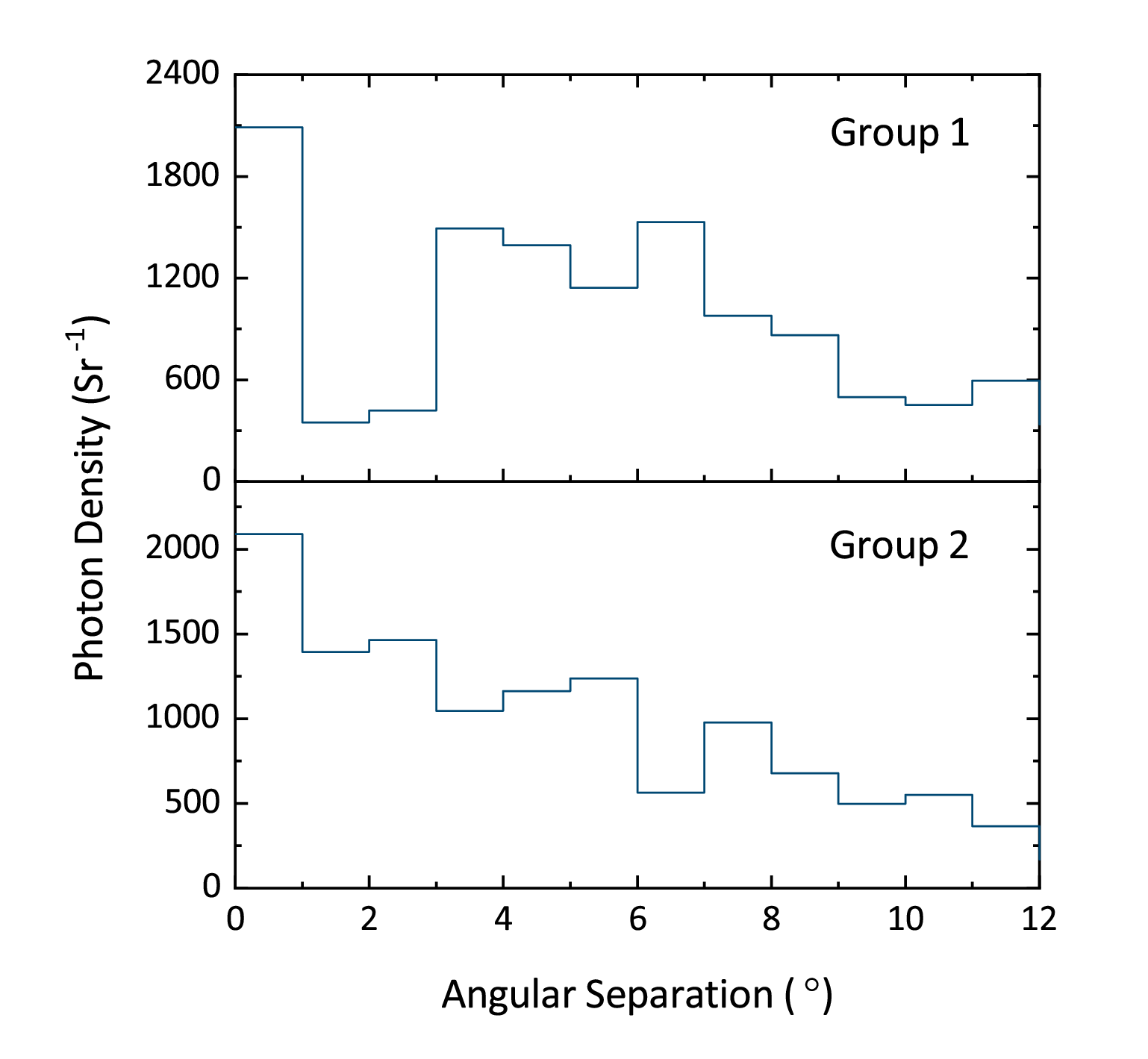}
		\caption{{The photon density-angle relation of GRB photon events within the time interval of $\tau < \mathrm{-20~s}$.}}
		\label{fig:method1}
	\end{center}
\end{figure}

The above discussion seems imprecise since the rising edge of the main-burst stage is also included in the time interval of $\tau < 0$.
Such concern can be dispelled when we do the same analysis with the time interval of $\tau < \mathrm{-20~s}$ and obtain similar results~(see  Figure~\ref{fig:method1}). It can be seen that the aforementioned convergence still exists.

We can not rule out the possibility that the photon detection
efficiency of Fermi detectors may have angle dependence so that the
observed angle dependence of the photon events presented in Figs.~1
and 2 is due the artificial consequence of the Fermi data but
not due to the realistic physical mechanism. Therefore the results
of our Figs.~1 and 2 can only serve as a hint to suggest the
possibility for the existence of the pre-burst stage, and we need
to perform more sophisticated study with detailed analysis on the
photon events about the pre-burst stage of GRBs.

\subsection{Classification of the pre-burst stage} \label{sec:2.2}
Amelino-Camelia et al.~\cite{method2} first suggested that the data from GRBs could be used to test the quantum gravity effect.
Many studies~(see, e.g., References~\cite{oldformula,Ellis:app,Rodri,Lamon:grg,DisCan,Abdo:1,Abdo:2,Xiao:2009xe,shaolijing,zhangshu,Xu:2016zxi,Xu:2016zsa,note-added,Xu:2018,Liu:2018,Ellis:1999sd,Chang:2012gq,Ellis:2018lca}) applied this method to search for the light speed variation caused by the Lorentz violation~(LV) effect or the cosmic matter effect.
Some of these studies~\cite{shaolijing,zhangshu,Xu:2016zsa,Xu:2016zxi,Xu:2018} suggest that some high-energy photons observed after time zero $t_{\mathrm{obs}}^0$ are emitted before $t_{\mathrm{in}}^0$ at source due to the light speed variation. Such effect is especially important for higher-energy~(about 10-300~GeV) photons.

We suppose that the pre-burst and main-burst photons can be distinguished via their geometric features on a certain coordinate system. Since the light speed variation is a primary cause of the difficulty to distinguish photons from two stages, the coordinate system should be able to demonstrate such speed variation effect. Luckily, some previous works~\cite{shaolijing,zhangshu,Xu:2016zxi,Xu:2016zsa,note-added,Xu:2018} proposed a coordinate system which meets our needs well, and detailed descriptions are given in the following.

The light speed variation might be caused by the Lorentz violation effect or by the cosmic medium effect. When the energy of a photon is much lower than the Planck scale $E_{\mathrm{Pl}}$, its
speed variation can be written in a general form of the leading
term of Taylor series~\cite{oldformula,Ellis:app,Rodri,Lamon:grg,DisCan,Abdo:1,Abdo:2,
	Xiao:2009xe,shaolijing,zhangshu,Xu:2016zxi,Xu:2016zsa,note-added,Xu:2018,Liu:2018}
\begin{equation}
v(E)=c\left[1-s_n \frac{n+1}{2}
\left(\frac{E}{E_{\mathrm{LV},n}}\right)^n\right],
\label{eq:2}
\end{equation}
where $s_n$ represents whether the photon speed is faster~($s_n=-1$) or slower~($s_n =1$) comparing to the light speed constant $c$ for $E\rightarrow 0$ and $E_{\mathrm{LV},n}$ represents a LV scale to be determined by data for the light speed variation, with $n=1$ corresponding to the simplest linear correction
\begin{equation}
v(E)=c\left(1-s\frac{E}{E_{\mathrm{LV}}}\right).
\label{eq:2b}
\end{equation}
where $s$ represents whether the photon speed is faster~($s =-1$) or slower~($s =1$) comparing to $c$ for $E\rightarrow 0$ with $E_{\mathrm{LV}}=E_{\mathrm{LV,1}}$.

Equation~(\ref{eq:2}) shows that the speed of a photon depends on its energy. So the photons simultaneously emitted at source can be observed at different times. The time difference $\Delta t_{\mathrm{LV}}$ caused by the speed variation is~\cite{oldformula,newformula}
\begin{equation}
{\Delta t_{\mathrm{LV}}} =({1+z})\frac{\kappa}{E_{\mathrm{LV}}},
\label{eq:3}
\end{equation}
where
\begin{equation}
\kappa  =s\frac{E_{\mathrm{obs}}-E_{\mathrm{peak}}}{H_0(1+z)}\int_{0}^{z} \frac{1+\zeta}{\sqrt{\Omega_m(1+\zeta)^3+\Omega_\Lambda}}d\zeta,
\label{eq:3b}
\end{equation}
in which $H_0=67.3\pm 1.2~\mathrm{kms^{-1}Mpc^{-1}}$ is the Hubble expansion constant; $[\Omega_m,\Omega_\Lambda]=[0.315_{-0.017}^{+0.016},0.685_{-0.016}^{+0.017}]$ are the cosmological constants~\cite{pgb} of pressureless matter density and dark energy density of the universe. $E_{\mathrm{peak}}$~(the energy of the first main peak photons, detected by GBM) can be omitted because it is negligible comparing to $E_{\mathrm{obs}}$~(detected by LAT). Since the redshifts of observed GRBs are of the same scale, $\kappa$ is roughly proportional to the observed energy $E_{\mathrm{obs}}$ of each photon.

Taking into account of the intrinsic emitting time $t_{\mathrm{in}}$ of the photon at the source, we obtain the observed time since $t^0_{\mathrm{obs}}=0$ as
\begin{equation}
t_{\mathrm{obs}}=\Delta t_{\mathrm{LV}}+t_{\mathrm{in}}({1+z}),
\label{eq:4a}
\end{equation}
from which we get
\begin{equation}
\tau= \frac{t_{\mathrm{obs}}}{1+z}=\frac{\Delta t_{\mathrm{LV}}}{1+z}+t_{\mathrm{in}} =\frac{\kappa}{E_{\mathrm{LV}}}+t_{\mathrm{in}}.
\label{eq:4b}
\end{equation}
Thus we arrive at a simple relation
\begin{equation}
\tau=\frac{\kappa}{E_{\mathrm{LV}}}+t_{\mathrm{in}}.
\label{eq:4}
\end{equation}
Equation~(\ref{eq:4}) allows us to construct a coordinate system whose horizontal axis is $\kappa$ and the vertical axis is $\tau$. This coordinate axis system demonstrates the speed variation effect because all simultaneously emitted photons will be plotted on a straight line, whose slope is $1/E_{\mathrm{LV}}$ and the intercept is $t_{\mathrm{in}}$, regardless of the energies and redshifts of these photons. In previous studies on high energy photons~\cite{Xiao:2009xe,shaolijing,zhangshu,Xu:2016zxi,Xu:2016zsa,note-added,Xu:2018,Liu:2018}, a regularity has been found that several high energy GRB events fall on a straight line in the  $\kappa$-$\tau$ plot to indicate a light speed variation. Such studies also suggest a scenario that
these high energy photons are emitted earlier than low energy photons at source. The purpose of the present paper is to re-check the $\kappa$-$\tau$ plot of high energy photons from a machine
learning analysis.

\begin{figure}
	\begin{center}
		\includegraphics[width=0.45\linewidth]{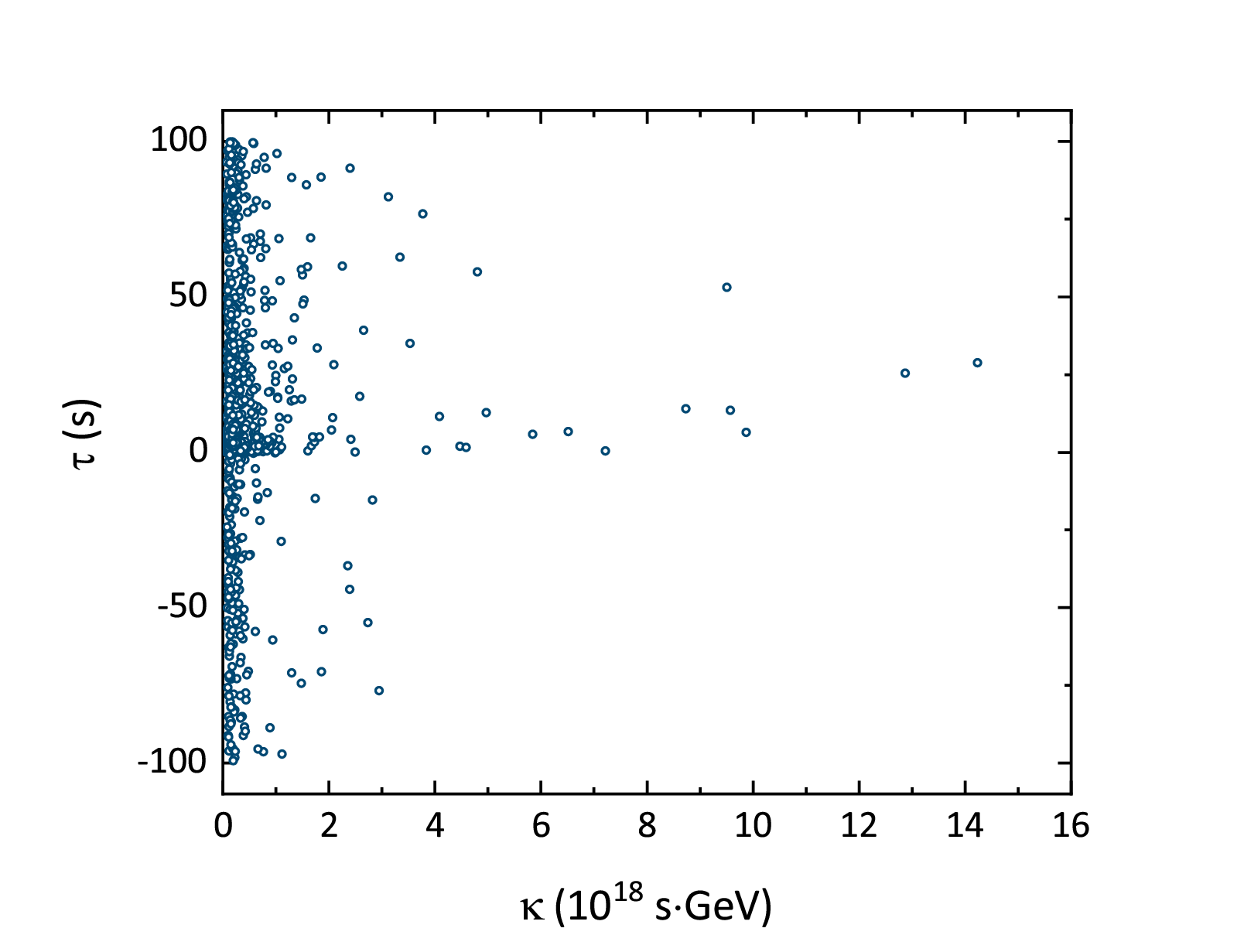}
		\vspace{0.2cm}
		\caption{{The $\tau$-$\kappa$ plot of Group~2.} Each blue circle represents a photon with intrinsic energy $E_{\mathrm{in}}>\mathrm{1~GeV}$. There are 979 photons in total.}
		\label{fig:3}
	\end{center}
\end{figure}

Since the speed variation effect is more significant for high-energy photons, we only take the higher-energy Group~2 into consideration, and the $\tau$-$\kappa$ plot is shown in Figure~\ref{fig:3}. The boundary of the pre-burst and main-burst stages can be roughly seen by naked eyes. To concretely illustrate such boundary, we apply a primary method of machine learning to produce classifications. Such machine learning method resembles the $K$-means method~\cite{k-means} with remarkable application in recent works~(see, e.g., Reference~\cite{N-body}).

The $K$-means classification method, first proposed by Macqueen,
is a process to classify a population of elements into $K$ clusters via
their geometric features. If $\Sigma=\{\Sigma_1,\Sigma_2,\cdots,S_K\}$ is a
classification of point set $V$, that is, any element of $V$
belongs to and only belongs to an $\Sigma_i\in S$. The purpose of the
$K$-means method is to find an optimal classification $\Sigma$
minimizing the error function
\begin{equation}
w(\Sigma)=\sum_{i=1}^{K}\left[\sum_{z\in \Sigma_i}d(z, u_i)^2\right],
\label{eq:5}
\end{equation}
where $z$ is elements of set $V$. In Macqueen's
work~\cite{k-means}, $u_i$ is the average coordinate of subset
$\Sigma_i$, and $d(z, u_i)$ is the Euclidean distance function.
Function $w(\Sigma)$ represents the divergence degree of classification $\Sigma$.

Finding the global optimal classification $\Sigma$ is computationally
difficult. However, the $K$-means method provides an efficient way
to produce a local optimal classification iteratively. The
$K$-means method first begins with $K$ random points
$\{u_1^0,u_2^0,\cdots,u_K^0\}$, then calculates $d(z, u_i^0)$ of
each $z\in V$, and places $z$ to the nearest subset $\Sigma_i^0$
according to the distance $d(z, u_i^0)$. The next step is to set
the average coordinate of each $\Sigma_i^0$ as the new point $u_i^1$.
Repeat the above process until the change of $\{u_i^n\}$ is
negligible. It is obvious that the $K$-means method gives a local
optimal classification $\Sigma$ because an arbitrary change of $\Sigma$ will
increase the error function $w(\Sigma)$ unless $\Sigma$ is unstable.

We develop a new method similar to the $K$-means method in our work.
The modifications are shown below.
\begin{enumerate}
	\item The dimension $N$ is set as $N=2$ because our coordinate axis system is $2$-dimensional.
	\item Since the simplest case of the K-means method is K=2, we set K=2 to
	illustrate the efficiency and enlightenment of the analysis.
	\item Instead of calculating the average coordinate of subset $\Sigma_i$ in each iteration as $u_i$,
	we compute the linear fitting line $\tau=k_i \kappa+b_i$ as $u_i$ of $\Sigma_i$, and $\{u_1^0,u_2^0,\cdots,u_K^0\}$ are $K$ random lines.
	\item We use the Euclidean distance function between a point $z(\kappa, \tau)$ and the line $u_i$ as $d(z, u_i)$:
	\begin{equation}
	d(z, u_i)=\frac{|k_i \kappa-\tau+b_i|}{\sqrt{k_i^2+1}}.
	\label{eq:7}
	\end{equation}
\end{enumerate}

One of the modifications is to change the $K$ initial points into $K$ initial lines, where $K = 2$ in our work. Since we calculate $K$ fitting lines in each step instead of $K$ mean values, we rename the new method as the $K$-lines method.

\begin{figure}
	\begin{center}
		\includegraphics[width=0.45\linewidth]{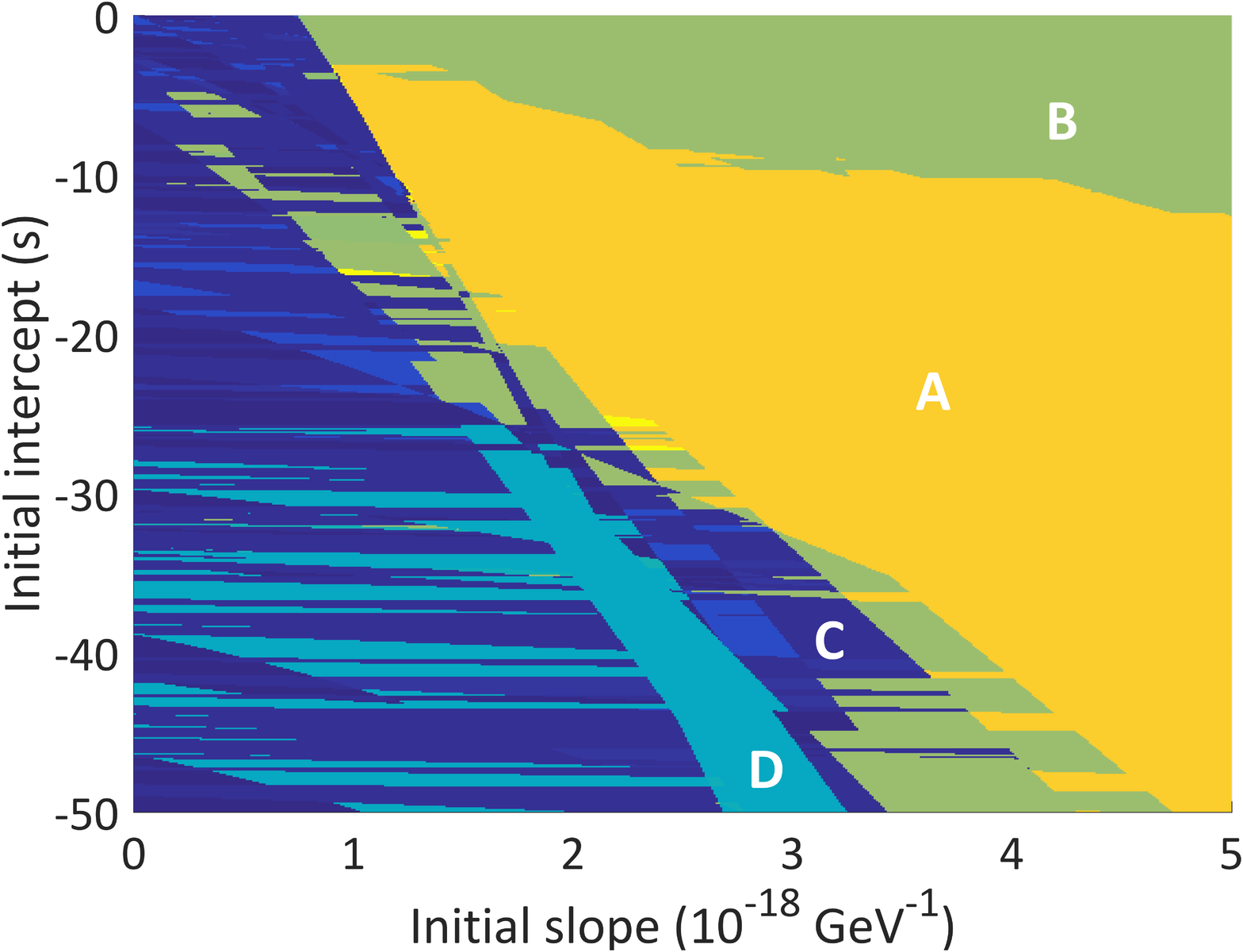}
		\vspace{0.3cm}
		\caption{{The phase diagram of results caused by different initial values.} Blocks A, B, C, D represent 4 main results caused by different initial values of the slope and the intercept of the second line.}
		\label{fig:4}
	\end{center}
\end{figure}

As an ideal situation, one may expect the $K$-lines method might be able to find a
global optimal classification regardless of the initially given
lines, however, in general situation there is still limitation of this machine learning method. From this sense,
the $K$-lines method is still not able to give the global optimal classification
and is initial-value dependent. Thus, we fix one
initial line as the $\tau$-axis, and perform a continuous scan over the parameter space of
another initial line to illustrate such initial-value dependence.~(see
Figure~\ref{fig:4}). It can be seen that even a
large range of initial values may result in a same classification,
and the 4 main classification results are represented by A, B, C,
D respectively. It is reasonable to expect the obtained classifications
in several options can provide a relatively complete view of the problem under study.

\begin{figure}
	\begin{center}
		\includegraphics[width=0.45\linewidth]{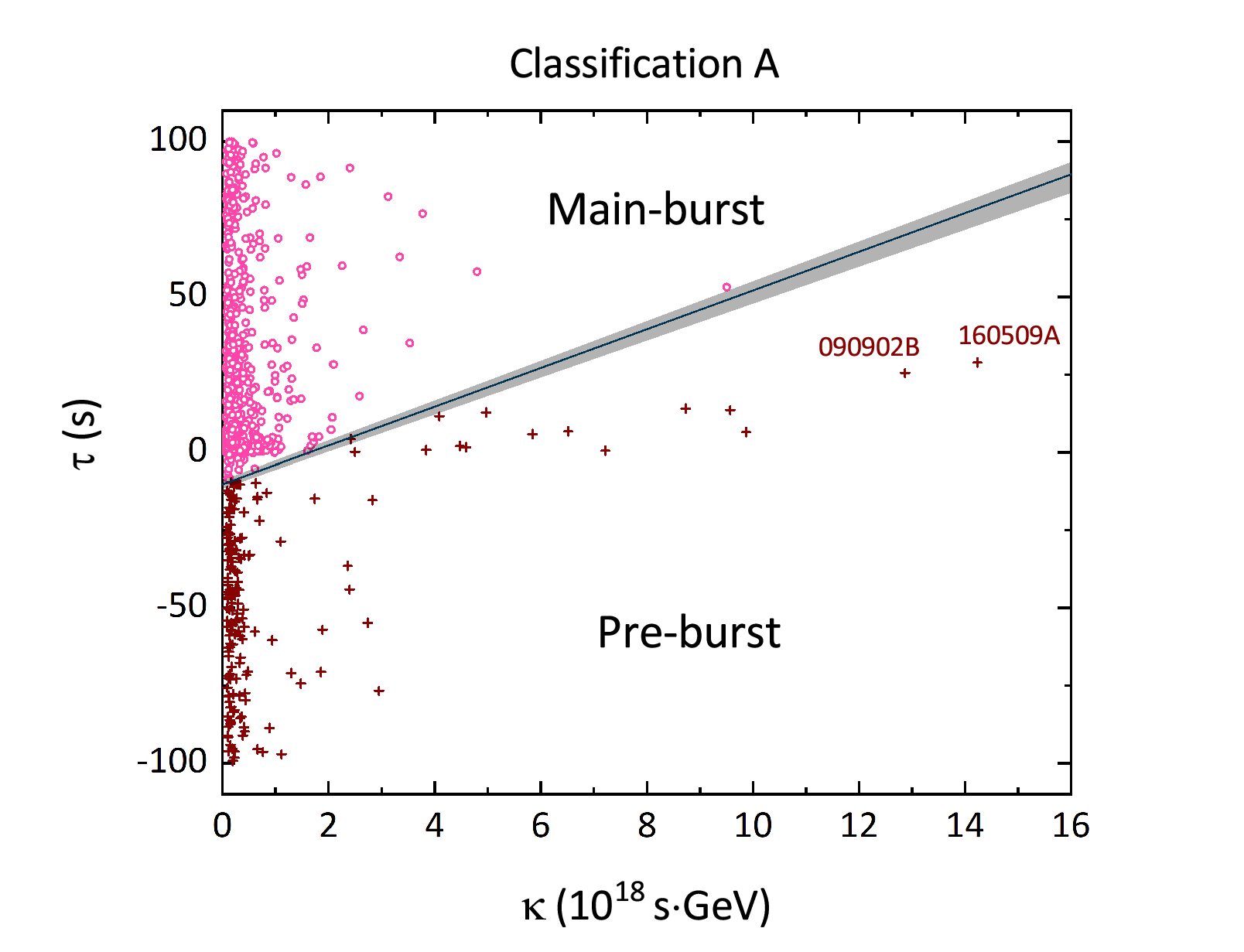}
		\includegraphics[width=0.45\linewidth]{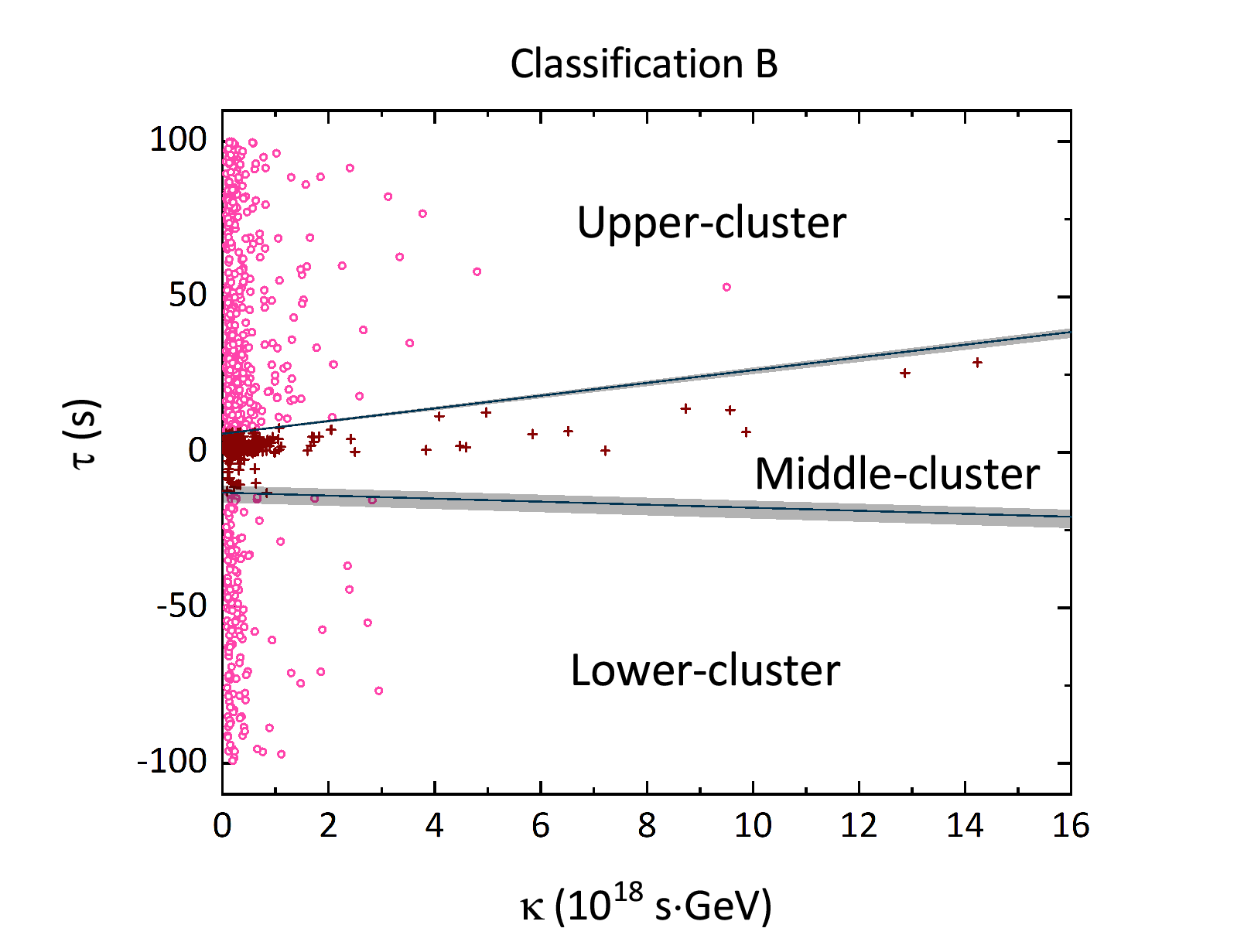}
		\caption{{Classifications A and B.} The boundary lines of each classification are colored in blue. In classification A, some high-energy photons of the pre-burst stage~(like the two marked photons from GRB~160509A and 090902B) are observed after time zero $t^0_{\mathrm{obs}}$ while they are emitted before $t^0_{\mathrm{in}}$ at source. The uncertainty of the boundary lines are shown as gray regions.
		}
		\label{fig:5all}
	\end{center}
\end{figure}

The largest block in Figure~\ref{fig:4} represents classification A, which provides a good referential classification of the pre-burst and main-burst stages~(see Figure~\ref{fig:5all}).
The two fitting lines of classification A read
\begin{align}
\tau &= \alpha_1\kappa+\beta_1, \nonumber \\
\tau &= \alpha_2\kappa-\beta_2,
\label{fig:add1c}
\end{align}
where
\begin{align}
[\alpha_1, \beta_1] &= [5.5 \pm 0.3, 22.7 \pm 0.1], \nonumber \\
[\alpha_2, \beta_2] &= [7.0 \pm 0.3, -52.3 \pm 0.2];
\label{fig:add1c2}
\end{align}
and the boundary line of classification A reads
\begin{equation}
\tau = \alpha_\mathrm{b}\kappa-\beta_\mathrm{b},
\label{fig:add1}
\end{equation}
where
\begin{equation}
[\alpha_\mathrm{b}, \beta_\mathrm{b}] = [6.2 \pm 0.2, -10 \pm 1],
\label{fig:add1-2}
\end{equation}
which is found by machine learning automatically. Considering the $10\%$ uncertainties of the energies of LAT photons~\cite{LATdata}, we produce the uncertainties of coefficients in the above equations (here we us the $68\%$ containment half width as the uncertainty).
It can be seen that some high-energy~(high-$\kappa$) photon events observed after time zero are classified by machine learning into the pre-burst stage. This result is interesting because it indicates that GRBs can emit considerably high-energy photon events which are belong to the pre-burst stage at source, and these high-energy photon events are detected after time zero due to the light speed variation effect.

Classification B is another classification also found by machine learning automatically. This result deserves notice because it divides all photons into three clusters~(see Figure~\ref{fig:5all}). The coefficients of the two fitting lines of classification B read
\begin{align}
[\alpha_1, \beta_1] &= [3.5 \pm 0.3, 16.9 \pm 0.1], \nonumber \\
[\alpha_2, \beta_2] &= [1.35 \pm 0.05, 0.87 \pm 0.02];
\label{fig:add1d}
\end{align}
and the coefficients of the boundary lines of classification B read
\begin{align}
[\alpha_\mathrm{b,1}, \beta_\mathrm{b,1}] &= [2.0 \pm 0.1, 6.0 \pm 0.4], \nonumber \\
[\alpha_\mathrm{b,2}, \beta_\mathrm{b,2}] &= [-0.49 \pm 0.02, -13 \pm 3];
\label{fig:add1b}
\end{align}
which are also produced by machine learning automatically.
By definition, the boundary lines of a classification result are the two angular bisectors of the two fitting lines, and the two angular bisectors are perpendicular to each other. In classification B, the two boundary lines separate all photons into three clusters, though the upper-cluster and lower-cluster of classification B are indeed fitted with the same fitting line. (The two boundary lines of classification B in Figure~\ref{fig:5all} do not look perpendicular because the axis is stretched).
However, the separation of the upper-cluster and lower-cluster implies that the geometric features of the middle-cluster are quite unique: the middle-cluster is likely the ``prompt phase'', since it contains a large number of high-energy photons; the lower-cluster is the ``pre-burst'' stage; and the upper-cluster is the ``afterglow''.

\begin{figure}
	\begin{center}
		\includegraphics[width=0.45\linewidth]{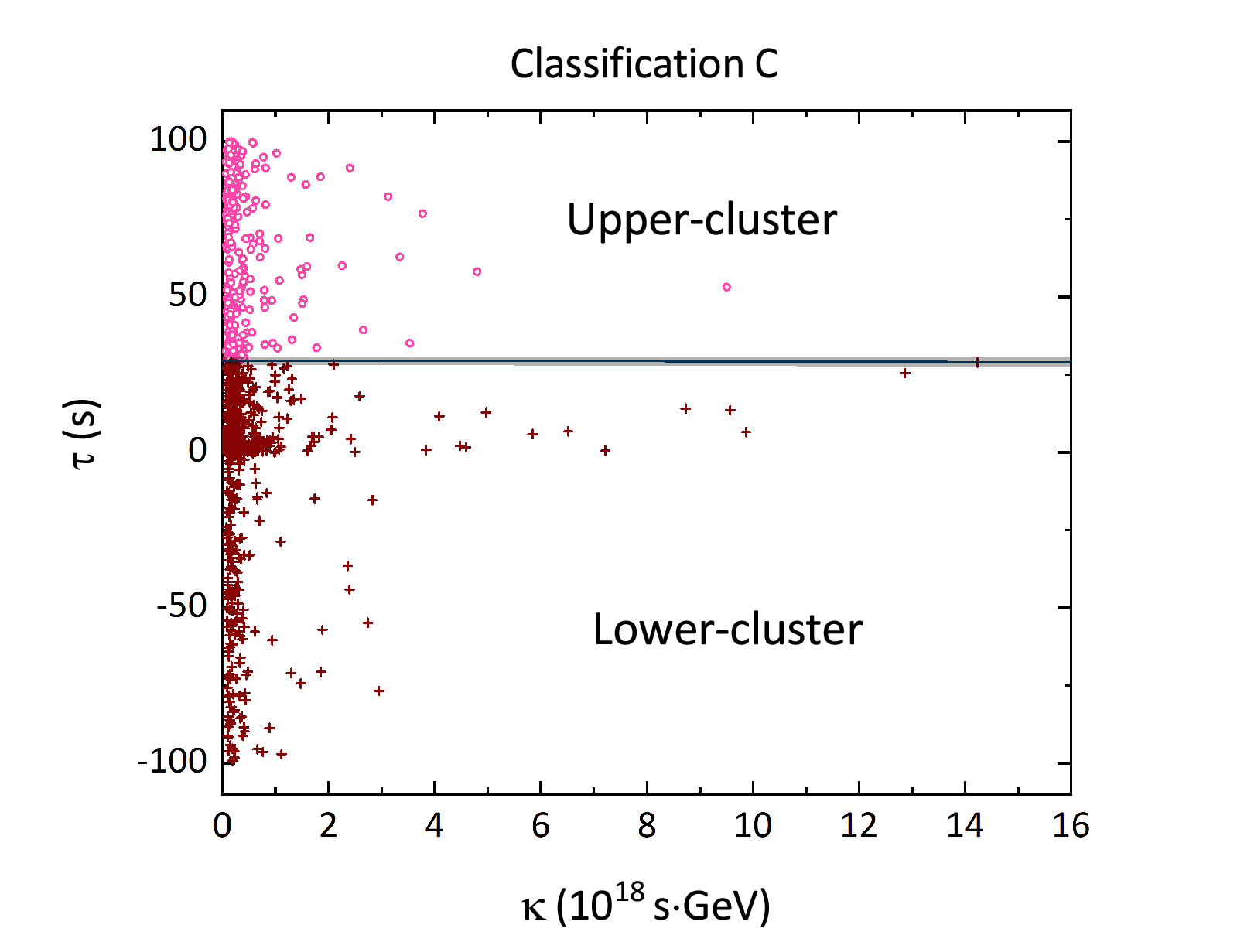}
		\includegraphics[width=0.45\linewidth]{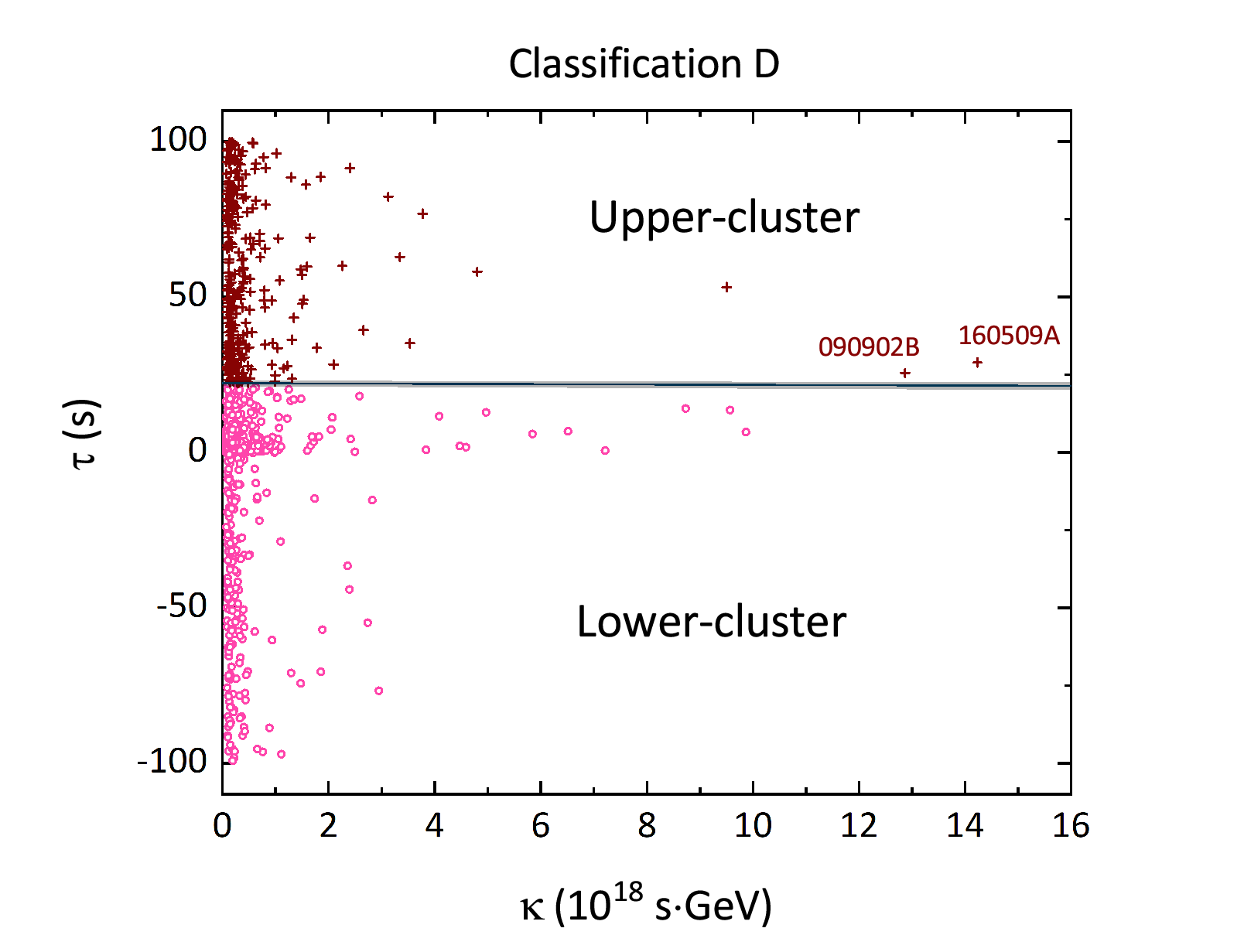}
		\caption{{Classifications C and D.} In classification D, two high-energy photon events of GRB 090902B and 160509A are separated from the lower-cluster. The uncertainty of the boundary lines are shown as gray regions.}
		\label{fig:append5}
	\end{center}
\end{figure}

Classification C is like a rough version of classification B~(see Figure~\ref{fig:append5}). It roughly combine the pre-burst stage~(the lower-cluster) and the prompt phase~(the middle-cluster) together.The coefficients of the two fitting lines of classification C read
\begin{align}
[\alpha_1, \beta_1] &= [-1.9 \pm 0.2, 68.1 \pm 0.1], \nonumber \\
[\alpha_2, \beta_2] &= [1.7 \pm 0.1, -5.42 \pm 0.03];
\label{eq:bound_ca}
\end{align}
and the coefficients of the boundary line of classification C read
\begin{equation}
[\alpha_\mathrm{b}, \beta_\mathrm{b}] = [-0.03 \pm 0.02, 30 \pm 1],
\label{eq:bound_c}
\end{equation}

\begin{figure}
	\begin{center}
		\includegraphics[width=0.45\linewidth]{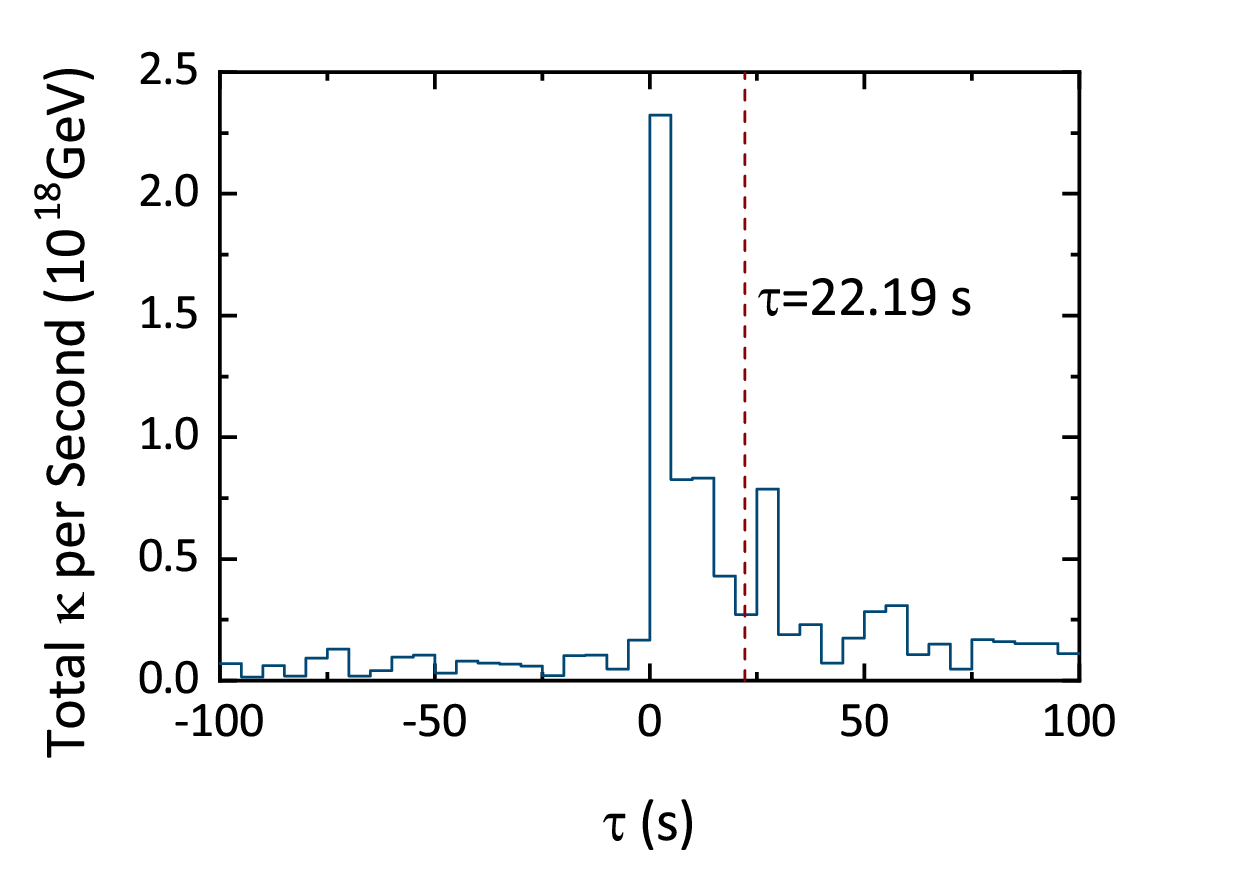}
		\caption{{The $\kappa$ density curve versus $\tau$ fo classification D.} The curve roughly represents the time distribution of the energy density, and the intercept of the boundary line of classification D ($\tau=\mathrm{22.19~s}$) locates in a local minimum point.
			\label{fig:kapa}
		}
	\end{center}
\end{figure}

Classification D is very close to classification C~(see Figure~\ref{fig:append5}). The coefficients of the two fitting lines of classification D read
\begin{align}
[\alpha_1, \beta_1] &= [1.4 \pm 0.1, -8.54 \pm 0.04], \nonumber \\
[\alpha_2, \beta_2] &= [-1.8 \pm 0.1, 58.78 \pm 0.04];
\label{eq:bound_da}
\end{align}
and the coefficients of the boundary line of classification D read
\begin{equation}
[\alpha_\mathrm{b}, \beta_\mathrm{b}] = [-0.06 \pm 0.02, 22 \pm 1],
\label{eq:bound_d}
\end{equation}
In fact, by plotting the $\kappa$ density curve of all photons in Group~2, this boundary line~(to be precise, the intercept $\tau=22.19$ of the boundary line) locates in a local minimum point of this curve~(see Figure~\ref{fig:kapa}).

\subsection{Analysis of classification results} \label{sec:2.3}
\begin{figure}
	\begin{center}
		\includegraphics[width=0.5\linewidth]{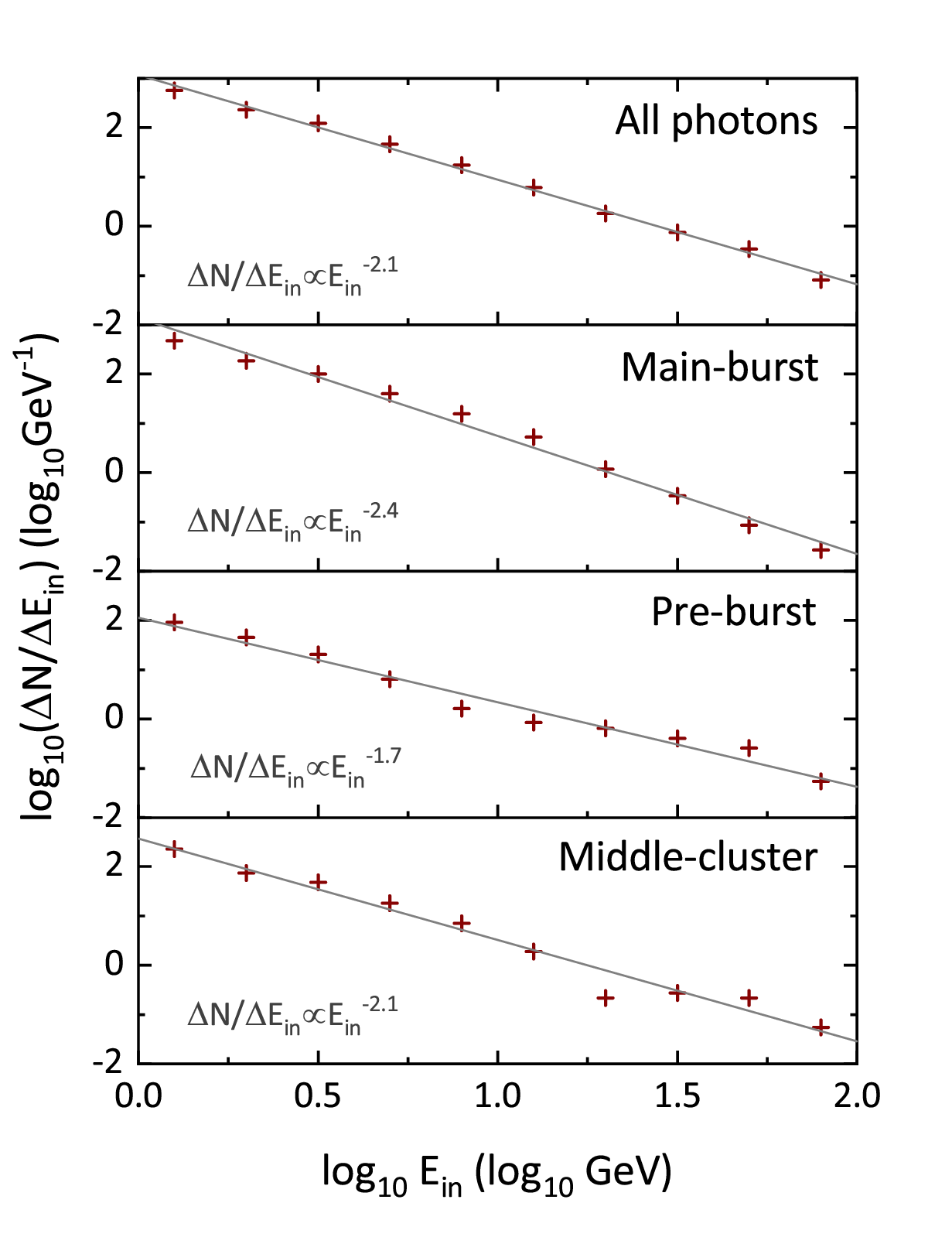}
		\caption{{The plots of intrinsic energy distributions.} In logarithmic coordinate systems, the power-law functions are shown in grey lines.}
		\label{fig:8}
	\end{center}
\end{figure}

We analyse the classification results based on above referential classifications.
The energy distributions of the pre/main-burst stages, the middle-cluster, and all photons are present in Figure~\ref{fig:8}. It can be seen that the energy spectrum of the pre-burst stage is more flat than the main-burst stage, which indicates that the pre-burst stage has a larger proportion of higher-energy photons in comparison to that of the main-burst stage.

We then follow Reference~\cite{Xu:2018} by adopting a general method with a complete scan over the $E_{\mathrm{LV}}$ space to find the light speed variation effect with the maximization of the $S(E_{\mathrm{LV}})$-function, which is defined as
\begin{equation}
S(E_{\mathrm{LV}})=\frac{1}{N-\rho}\sum_{i=1}^{N-\rho}\ln\left(\frac{\rho}{t_{i+\rho}-t_i}\right),
\label{eq:8}
\end{equation}
where $E_{\mathrm{LV}}$ is the LV scale~(mentioned in Equation~(\ref{eq:2})); $t_i$ is the intrinsic emitting time~(mentioned in Equation~(\ref{eq:4})) of the $i$-th photon in the data set~(sorted in ascending order); $N$ is the size of the data set; $\rho$ is a pre-set parameter~
(the recommended value of $\rho$ in Reference~\cite{Xu:2018} is $\rho=5$, with $\rho=1$ to $6$ also examined to give similar results).
Due to the light speed variation effect, a light curve with peak at source will be flattened when observed. By adopting the aforementioned method, we can identify the undetermined $E_{\mathrm{LV}}$ of the light speed variation as the location of the maximized $S(E_{\mathrm{LV}})$-function. The $S(E_{\mathrm{LV}})$-function can be traced back to the concept of ``information entropy'' based on the expectation that a trial of a true $E_{\mathrm{LV}}$ can produce a maximal value of the $S(E_{\mathrm{LV}})$-function. It can be seen that a pair of simultaneously emitted photons is able to create a large contribution to the $S(E_{\mathrm{LV}})$-function, if $E_{\mathrm{LV}}$ corresponds to the correct LV scale. As a result, the location of the peak in the $S(E_{\mathrm{LV}})$ plot represents the undetermined $E_{\mathrm{LV}}$ of the light speed variation.

Without loss of generality, we apply such method only on photons within the energy interval of $E_{\mathrm{in}}>\mathrm{20~GeV}$~(these photons belong to set~3 in Reference~\cite{Xu:2018}), and provide the $\sigma$-regions by the Monte-Carlo method, and the randomization process of the Monte-Carlo method is as follows.

\begin{enumerate}
	\item The size of each random data set is the same as each actual data set.
	\item The $\tau$ values are randomly permutated and then used as those of the random data sets.
	\item The intrinsic energies of each random data set are produced to satisfy the power-law functions mentioned earlier in Figure~\ref{fig:8}, and have the same range with the actual data set.
\end{enumerate}

\begin{figure}
	\begin{center}
		\includegraphics[width=0.45\linewidth]{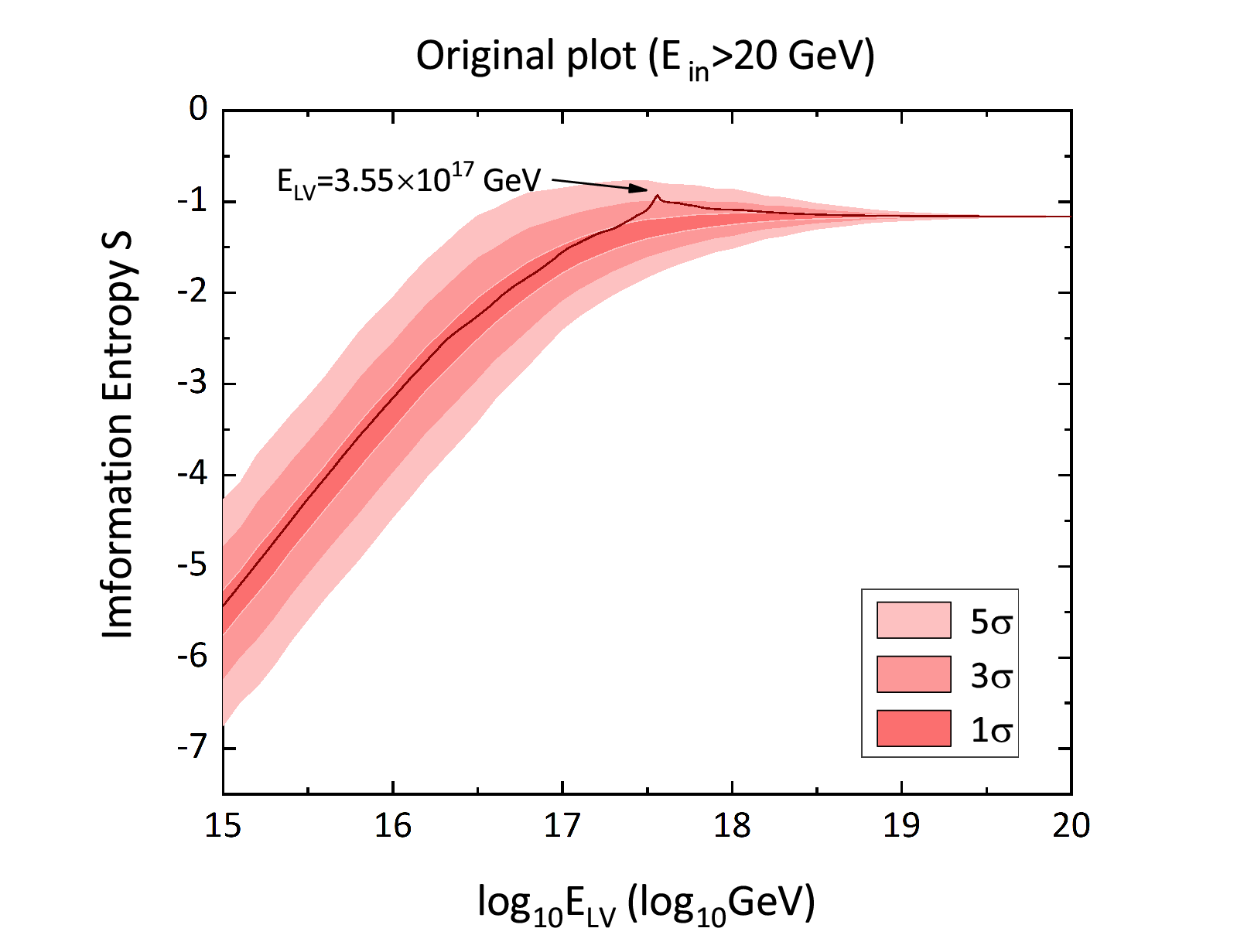}
		\includegraphics[width=0.45\linewidth]{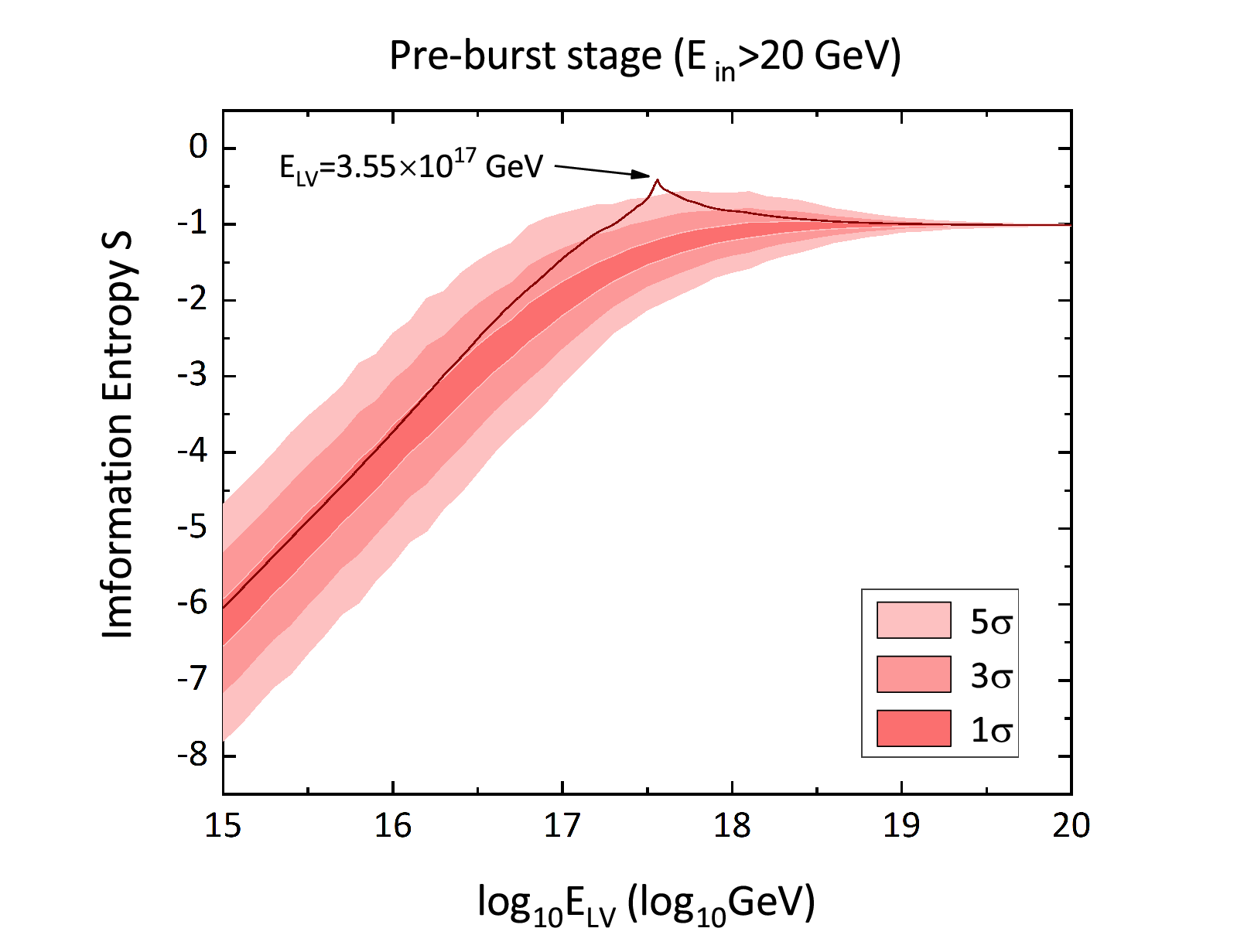}
		\includegraphics[width=0.45\linewidth]{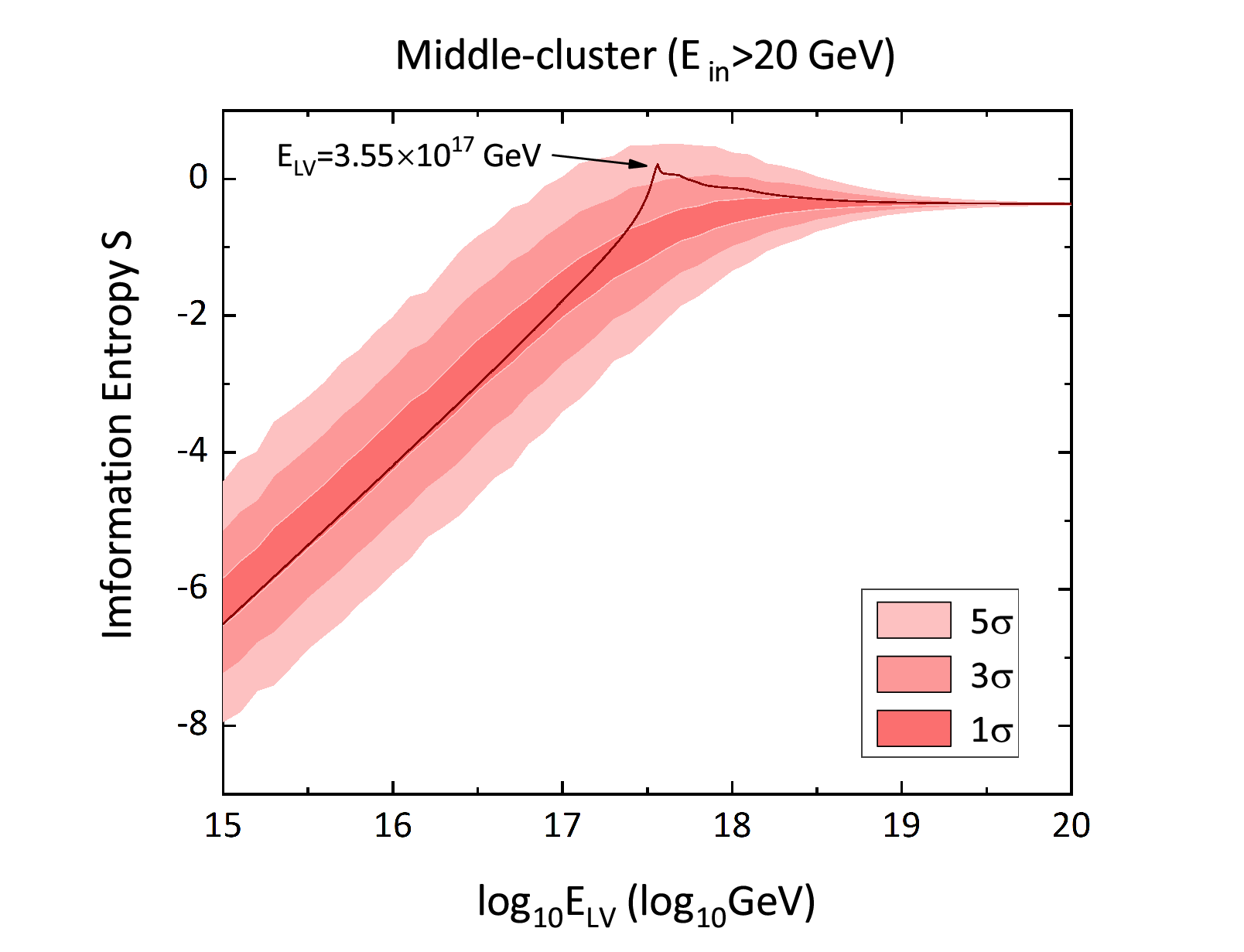}
		\caption{{The $S(E_{\mathrm{LV}})$ plots of 3 situations.} (a) is the original plot of the unclassified situation; (b) is the plot of the pre-burst stage in classification A; (c) is the plot of the middle-cluster in classification B.}
		\label{fig:6}
	\end{center}
\end{figure}

As shown in Figure~\ref{fig:6}, the $S(E_{\mathrm{LV}})$ plots of the pre-burst stage and the middle-cluster have sharper peaks in comparison to the original plot, and the plot of the pre-burst stage even exceeds the $5\sigma$-region at $E_{\mathrm{LV}}=\mathrm{3.55\times 10^{17}~GeV}$. This result also fits well with References~\cite{Xu:2016zsa,Xu:2016zxi,Xu:2018}.

\begin{figure}
	\begin{center}
		\includegraphics[width=0.45\linewidth]{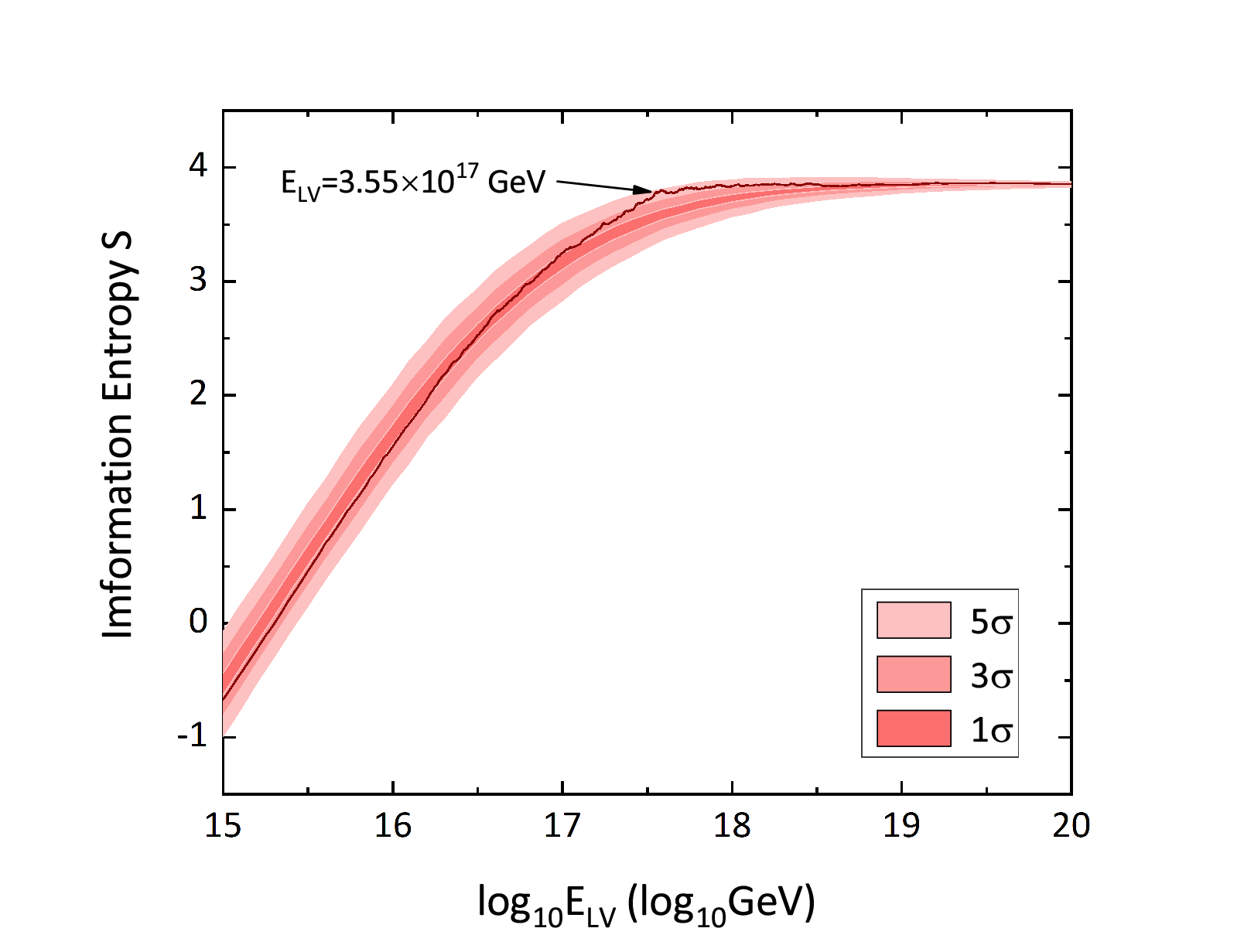}
		\caption{{The $S(E_{\mathrm{LV}})$ plot of the middle-cluster of classification B.} The energy interval is set as $E_{\mathrm{in}}>\mathrm{1~GeV}$, and the peak is also located at $E_{\mathrm{LV}}=\mathrm{3.55\times 10^{17}~GeV}$.
			\label{fig:sb}
		}
	\end{center}
\end{figure}

The middle-cluster of classification B roughly represents the prompt phase. Since this cluster has many high-energy photons, we can find a peak of the $S(E_{\mathrm{LV}})$ plot even if the energy threshold is considerably low~(see Figure~\ref{fig:sb}).

\begin{figure}
	\begin{center}
		\includegraphics[width=0.45\linewidth]{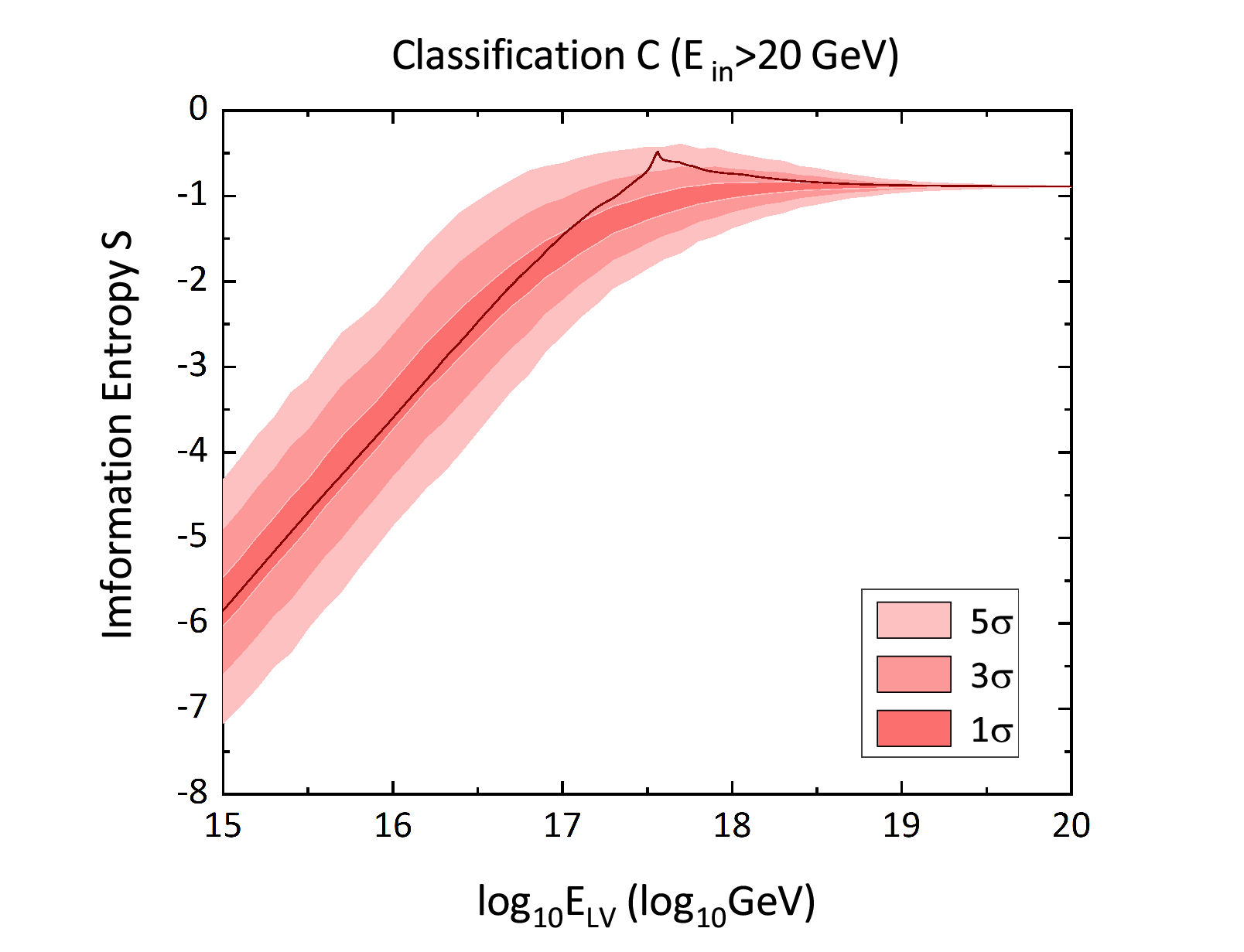}
		\includegraphics[width=0.45\linewidth]{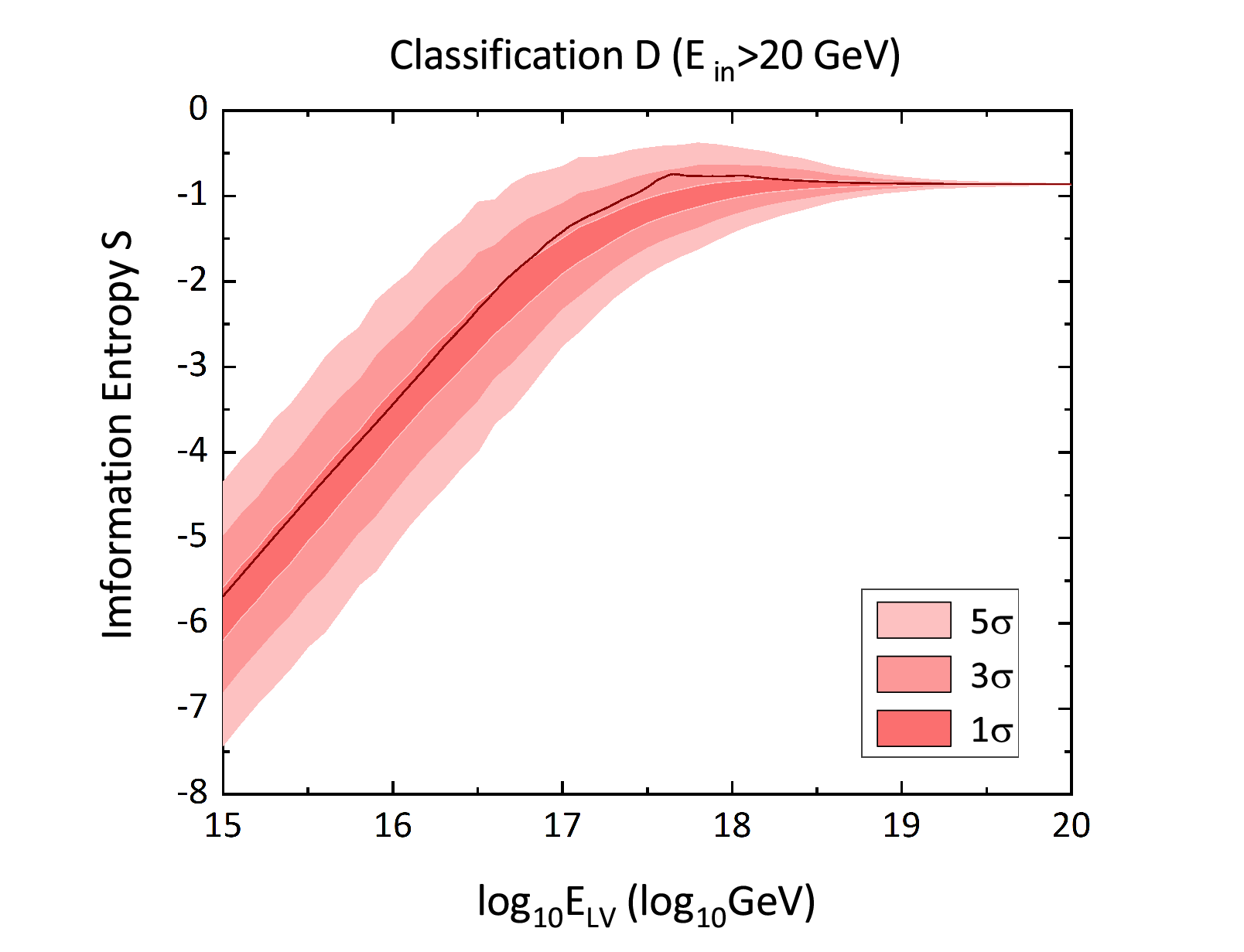}
		\caption{{The $S(E_{\mathrm{LV}})$ plots of the lower-clusters of classifications C and D.}
			\label{fig:s}
		}
	\end{center}
\end{figure}

We also provide the $S(E_{\mathrm{LV}})$ plots of the lower-clusters of classifications C and D~(see Figure~\ref{fig:s}). It can be seen that the plot of classification C has a peak,
which is compatible with those from classifications A and B.

Classification D also has a peak in the $S(E_{\mathrm{LV}})$ plot but the peak is flat. However, such flatness is not in conflict with the speed variation.
In fact, a flat distribution of light curve at source cannot produce a sharp peak for the observed $S(E_{\mathrm{LV}})$ plot.

\section{Discussion and Conclusion} \label{sec:4}
As extremely energetic events in the universe, GRBs provide us a powerful tool to analyse cosmic LV effect. By measuring time delay of high energy photons, previous works set lower limits for the LV energy scale. The study on GRB 090510 \cite{Abdo:2} reported a lower bound $E_{\rm LV}>1.2~E_{\rm Pl}$, and Ref.~\cite{Xiao:2009xe} also suggested a lower limit of $E_{\rm LV}>6.3~E_{\rm Pl}$. These works all assume that the high energy photons are from the same origin of the other photons, i.e., high energy photons cannot be intrinsically emitted earlier than the low energy photons. However, previous works~\cite{shaolijing,zhangshu,Xu:2016zsa,Xu:2016zxi,Xu:2018} suggested that some photons of GRBs may be emitted before the main-burst stage at source. In this work, we propose the pre-burst stage of GRBs by a first glimpse of the FGST data. Considering the light speed variation effect, we adopt a primary method of machine learning to analyse the high-energy photon events of 25 bright GRBs with known redshifts. The machine learning method automatically provides 4 main classifications, which propose the existence of the pre-burst stage of GRBs in all cases.  Classification A indicates that some high-energy photons observed after the onset of the prompt low energy photons are emitted in the pre-burst stage before the prompt phase at source. Classification B suggests a special middle-cluster which includes most of the high-energy photon events. We then adopt a general method to extract the light speed variation from the pre-burst stage of classification A and the middle-cluster of classification B, and get consistent result of a light speed variation at $E_{\mathrm{LV}}=\mathrm{3.55\times 10^{17}~GeV}$.
Such speed variation conversely confirms the existence of the hitherto unknown pre-burst stage of GRBs with earlier emission of both higher-energy and low energy photons at source. Most recently, a direct observation of pre-burst events of GeV scale photons was reported in Ref.~\cite{Zhu:2021}, and such observation serves as a strong support of our proposal of the pre-burst stage of GRBs from the machine learning method.

The novel pre-burst stage of GRBs can also provides novel insights about the dynamics of GRBs.
GRBs are likely to have extremely high temperature initially followed by a cooling process, which is an analogue of the big bang. A large amount of high energy photons are emitted during this phase until there is an intense explosion, which leads to the prompt phase. Following the prompt phase, the temperature of GRBs continues to decrease for several weeks to a year, forming the afterglow. The cooling fact of our pre-burst model explains why high energy photons are emitted prior to lower energy photons. Interestingly, all of the three stages can be automatically revealed by the $K$-lines method.
We thus conclude that the machine learning method provides a powerful tool to analyse the GRB data with fruitful results.

\bibliographystyle{unsrt}


\end{document}